%% file: main.tex
\documentclass[numbers]{article}
\usepackage{arxiv}

\usepackage[utf8]{inputenc} 
\usepackage[T1]{fontenc}    
\usepackage{hyperref}       
\usepackage{url}            
\usepackage{booktabs}       
\usepackage{amsfonts}       
\usepackage{nicefrac}       
\usepackage{microtype}      
\usepackage{amsmath}         
\usepackage{cleveref}       
\usepackage{lipsum}         
\usepackage{graphicx}
\usepackage[font=footnotesize]{caption}
\usepackage{lineno}
\usepackage[numbers,sort&compress]{natbib}

\usepackage{doi}
\usepackage[separate-uncertainty=true]{siunitx}    

\usepackage[font=scriptsize, labelfont=bf]{caption}

\title{One size does not fit all: session-specific effects of deep brain stimulation on hand-motor kinematics in Parkinson's disease}

\usepackage{authblk}

\newbox{\orcid}\sbox{\orcid}{\includegraphics[scale=0.06]{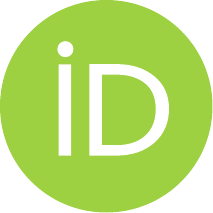}} 
\author[1,2,3]{%
	\href{https://orcid.org/0009-0003-1477-4912}{\usebox{\orcid}\hspace{1mm}Matthias Dold\thanks{\texttt{matthias.dold@donders.ru.nl}}}%
}
\author[3, 4, 1]{%
	\href{https://orcid.org/0000-0002-2032-8981}{\usebox{\orcid}\hspace{1mm}Joana Pereira\setcounter{footnote}{0}}%
}
\author[3, 4]{%
	\href{https://orcid.org/0000-0002-1703-6283}{\usebox{\orcid}\hspace{1mm}Volker A. Coenen}%
}
\author[3]{%
	\href{https://orcid.org/0000-0002-2836-0046}{\usebox{\orcid}\hspace{1mm}Bastian Sajonz}%
}
\author[5]{%
	\href{https://orcid.org/0009-0003-0594-7893}{\usebox{\orcid}\hspace{1mm}Sophia Gimple}%
}
\author[5, 7]{%
	\href{https://orcid.org/0000-0002-3589-5604}{\usebox{\orcid}\hspace{1mm}Yasin Temel}%
}
\author[2]{%
	\href{https://orcid.org/0000-0003-0030-7224}{\usebox{\orcid}\hspace{1mm}Marcus L.F. Janssen}%
}
\author[1]{%
	\href{https://orcid.org/0000-0001-6729-0290}{\usebox{\orcid}\hspace{1mm}Michael Tangermann\setcounter{footnote}{0}\thanks{\texttt{michael.tangermann@donders.ru.nl}}}%
}

\affil[1]{Data-Driven Neurotechnology Lab, Donders Institute for Brain, Cognition and Behaviour, Radboud University, Nijmegen, The Netherlands}
\affil[2]{Department of Clinical Neurophysiology, Maastricht University Medical Center, The Netherlands;  Mental Health and Neuroscience Research Institute, Maastricht University, The Netherlands}
\affil[3]{Department of Stereotactic and Functional Neurosurgery, University of Freiburg - Medical Center, Breisacher Straße 64, 79106, Freiburg, Germany}
\affil[4]{Brain-Links Brain-Tools, Intelligent Machine-Brain Interfacing Technology (IMBIT), University of Freiburg, Georges-Köhler-Allee 201, 79110, Freiburg, Germany
}
\affil[5]{Department of Neurosurgery, Maastricht University Medical Center, The Netherlands; Mental Health and Neuroscience Research Institute, Faculty Health, Medicine and Life Sciences, Maastricht University, The Netherlands}
\affil[6]{Department of Diagnostic and Interventional Radiology, University of Freiburg - Medical Center, Killianstrasse 5a, 79106, Freiburg, Germany}
\affil[7]{Istanbul Atlas University, Faculty of Medicine, Istanbul, Turkiye}

\renewcommand{\shorttitle}{Individual DBS modulation of hand-motor features}

\hypersetup{
pdftitle={One size does not fit all: session-specific effects of deep brain stimulation on hand-motor kinematics in Parkinson's disease},
pdfsubject={q-bio.NC, q-bio.QM},
pdfauthor={Matthias Dold, Joana Pereira, Volker A. Coenen, Bastian Sajonz, Peter Reinacher, Thomas Prokop, Sophia Gimple, Yasin Temel, Marcus L.F. Janssen, Michael Tangermann},
pdfkeywords={DBS, EEG, ECoG, neural decoding, Parkinson's Disease},
}

\newcommand{\handmotor}{hand-motor }

\newcommand{\pval}{$\textit{P}_{\text{BH}}$}

\begin{document}\maketitle

\begin{abstract}

Deep brain stimulation (DBS) is an effective treatment of motor symptoms in Parkinson's disease (PD), but it is unclear whether it changes fine hand movements in the same way for every patient. Increasingly, the effects of DBS are investigated based on symptom-relevant behavioral tasks, such as the CopyDraw task~\cite{Castano2019}. These tasks are designed to circumvent the limitations of standardized clinical assessments, which can be subjective, lack granularity, and lack robust test-retest reliability. 

We analyzed 35 sessions of 19 patients from a multi-center cohort (14 with externalized and 5 with fully implanted systems). Besides standard kinematics (mean velocity, acceleration, jerk, and drawing accuracy), which have been reported in a small sample before, we derived higher distributional moments and non-linear complexity measures, for 38 features in total. DBS effects were quantified with linear mixed models at the group level and with residualized regression and a random-effects meta-analysis per session. The DBS state was decoded per session with regularized linear discriminant analysis comparing the resulting decoding performance of different feature sets, with the objective of maximizing the session-specific utility of these features as labels to derive future physiomarkers for adaptive DBS.

DBS modulated all 38 features at the group level, most strongly velocity related features. Yet the group effect is insufficient to uncover large per session differences. We found that for every feature, the between-subject variability of the DBS effect exceeded the group-level effect. 
Only one feature showed a significant mean effect across sessions (detrended fluctuation analysis of the distance to the template), and it does not belong to the standard kinematic features used within the CopyDraw task before.
In six sessions without any significant modulation of standard features, complexity measures were modulated instead. For decoding, standard features alone were the best choice in ten of 35 sessions, but combining all 38 features yielded the highest average area under the receiver operating characteristic curve (0.700 vs.~0.677), with the largest gains in sessions that were hard to decode from standard features.

DBS effects on \handmotor behavior are individual. Deriving neural markers for aDBS from behavior therefore requires session-specific analyses considering individualized features beyond standard kinematics and group-level effects.

\end{abstract}

\keywords{DBS \and aDBS \and CopyDraw \and hand motor}

\include{introduction}

\include{methods}

\include{results}

\include{discussion}

\include{conclusion}

\include{acknowledgements}

\bibliographystyle{vancouver}   
\bibliography{bibliography} 

\end{document}

%% file: introduction.tex
\section{Introduction}

Parkinson's disease (PD) is a progressive neurodegenerative disorder~\cite{Bloem2021} caused by the loss of dopaminergic cells in the substantia nigra pars compacta. PD affects over 10 million people worldwide~\cite{Su2025}, with a rapid rise in prevalence over the last 20 years~\cite{Zhu2024}. Its characteristic motor symptom is bradykinesia, in combination with rigidity, resting tremor, and/or postural instability~\cite{Postuma2015}. These symptoms are linked to a broad range of motor deficits affecting the entire body~\cite{Moustafa2016}. Fine motor execution of the hand is one such domain that is heavily impacted by the disease. These impairments significantly restrict patients’ independence in activities of daily living~\cite{Foki2016}, which in turn negatively impact overall quality of life~\cite{WongYu2022}.

Because of the highly sensitive nature of fine motor control, handwriting and continuous drawing capabilities were found to be impaired already at a prodromal stage~\cite{Fereshtehnejad2019}. As such, \handmotor control is used for the early identification of PD and for the comparison of PD vs.~healthy controls~\cite{Aouraghe2023}. Two prominent clinical examples are the assessment of micro- or dysgraphia~\cite{Letanneux2014} and the Archimedes spiral drawing task~\cite{Luciano2016, Pullman1998}. Micrographia, in its progressive form, describes an amplitude decay observed during a continuous writing task. Similar continuous, rhythmic movements are analyzed during the spiral drawing task, which captures spatial trajectories, motor arrest, and tremor frequency~\cite{Wang2025}. Consequently, both tasks have been used as behavioral proxies to capture symptom severity and to monitor disease progression in interventional studies~\cite{Asci2025, Radmard2021, Zham2019, Riosurrego2019}. Such interventions primarily comprise dopamine replacement therapies~\cite{Fox2018} and/or deep brain stimulation (DBS)~\cite{deuschl:2006, Hartmann2019, Foote2025}. 
The comparison of treatment conditions in the \handmotor context (e.g., medication ON/OFF and DBS ON/OFF) mostly focuses on spatial and temporal kinematics---such as velocity, acceleration, and fluency---enabled by the use of digitizing tablets~\cite{Thomas2017}. Additionally, measures of spatial complexity and dynamic irregularities have been analyzed~\cite{Riosurrego2019, Wang2024}.

Our group has developed a computerized copy-drawing paradigm (the CopyDraw task)~\cite{Castano2019}, which requires participants to draw a predefined trace (template) under time pressure. The task instruction is to copy the target template ``as fast and as accurately as possible'', which adds a critical performance trade-off dimension to the task in the form of a speed-accuracy trade-off. 

Crucially, the CopyDraw task has been used to derive objective, continuous hand-motor performance labels. These behavioral labels enable the subsequent derivation of neural marker candidates from concurrent brain signals for adaptive deep brain stimulation (aDBS), which we recently demonstrated in a proof-of-concept utilizing invasive electrocorticography (ECoG) signals~\cite{dold2025}.

So far, the CopyDraw task has been investigated primarily with respect to its effectiveness~\cite{Castano2019, Castano2020} and utility~\cite{Dold2026} to derive neural markers for aDBS. With this work, we provide a comprehensive in-depth analysis of DBS effects on the behavioral, i.e., hand-motor, performance of patients with PD executing the CopyDraw task. Specifically, we address 1) how individual features are modulated under DBS using linear mixed models, and 2) whether we can optimize the DBS decoding from \handmotor behavior by including features for higher moments of the distributions and non-linear measures of complexity.

%% file: methods.tex
\section{Methods}
This section first describes the patient cohort and the CopyDraw task (Sections~\ref{sec:met_population}--\ref{sec:met_setup}) and the extraction of 38 \handmotor features grouped into \textit{standard}, \textit{distributional}, and \textit{complexity} sets (Section~\ref{sec:met_features}). We then detail the
statistical models used to quantify the DBS effect on these features at the group and the session level (Section~\ref{sec:met_dbs_effect}), and the pipeline for decoding the DBS state from each feature set (Section~\ref{sec:met_decoding}).

\subsection{Population}
\label{sec:met_population}
The data set comprises 35 sessions of 19 patients with PD performing the CopyDraw task~\cite{Castano2019} with DBS electrodes implanted. The full data set is publicly available~\cite{dold2026dataset}. For a comprehensive description we refer the reader to Dold et al.~\cite{Dold2026}.

\subsection{Experimental setup}
\label{sec:met_setup}
\begin{figure}
    \centering   
    \includegraphics[width=\linewidth]{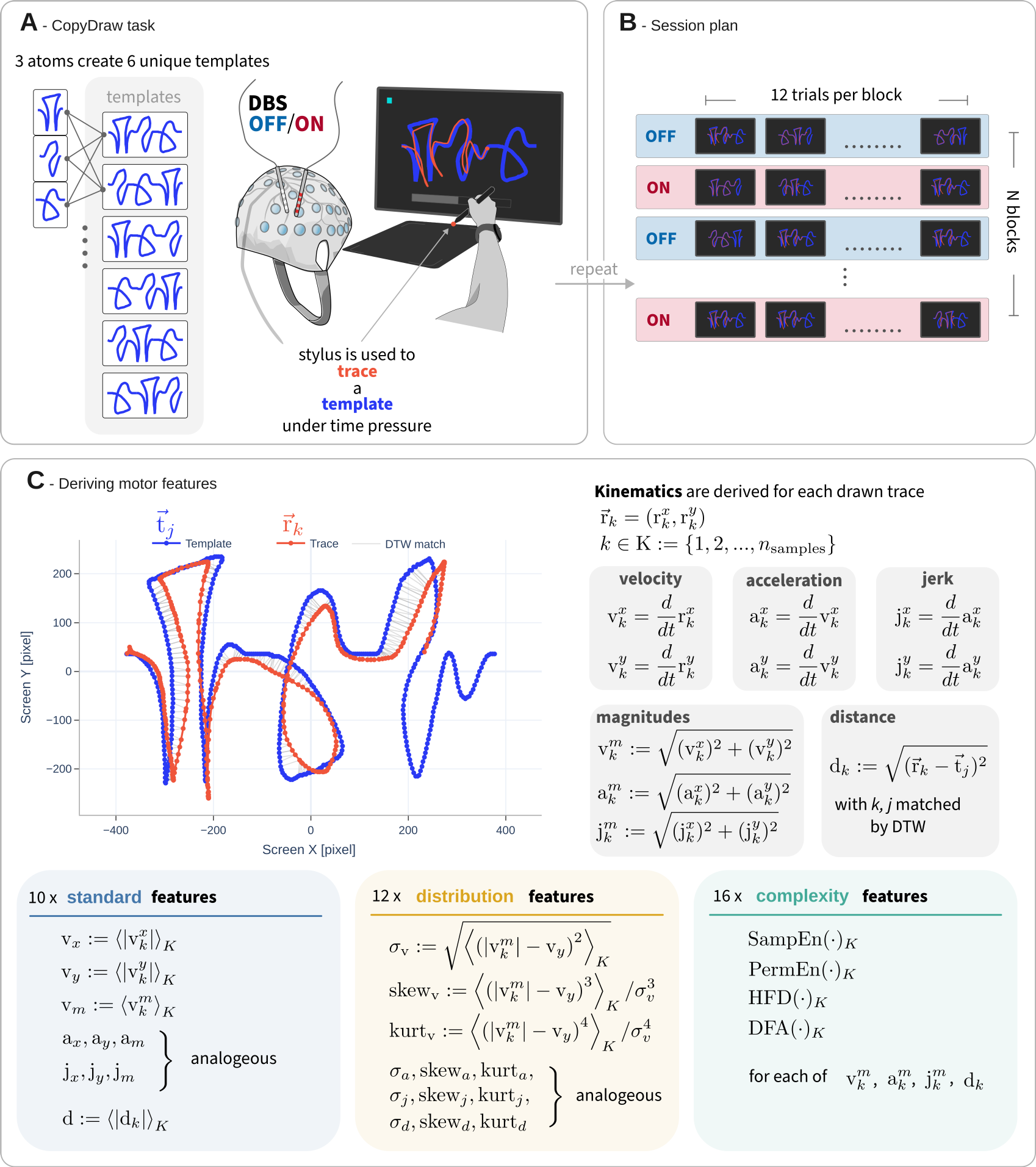}      
    \caption{Experimental setup of the CopyDraw experiments. \textbf{A}: During the CopyDraw task participants are asked to copy-draw a given template under time pressure. A digital stylus is used to control a cursor on a computer screen for this purpose. The target template (blue) and the current drawing (red) are displayed, and a grey bar at the bottom of the screen indicates the remaining time. \textbf{B}: Individual trials are repeated in blocks of 12 trials each. The DBS is alternated between ON and OFF for each of multiple blocks. \textbf{C}: For each drawing, i.e., trial, kinematics are derived from the sampled cursor position, a.k.a.\ the trace ($\vec{\text{r}}_k$, with k being the sample index out of the index set $K:=\{1,2,...,\text{n}_{\text{samples}}\}$). The derivative operator ($d/dt$) is here to be understood as a numeric differentiation using convolution with a modified sinc kernel. Additionally, dynamic time warping (DTW) is used to match trace ($\vec{\text{r}}_k$) and template ($\vec{\text{t}}_j$) points. The Euclidean distance between matched trace and template points defines the distance ($\text{d}_k$) for each trace point $k$. Three categories of scalar features derived from the kinematics and distances of individual trials are defined: 1) \textit{standard} features containing mean absolute velocity, acceleration, and jerk in x- and y-direction and in magnitude as well as mean distance between trace and template, 2) standard \textit{distributional} features containing standard deviation ($\sigma$), skew, and kurtosis of the magnitudes for velocity, acceleration, jerk and distances, and 3) \textit{complexity} features containing sample entropy (SampEn), permutation entropy (PermEn), Higuchi's fractal dimension (HFD)~\cite{Higuchi1988} and detrended fluctuation analysis (DFA) for the magnitudes of velocity, acceleration, jerk and the distances. }
    \label{fig:setup}
\end{figure}

During the CopyDraw task~\cite{Castano2019}, participants were asked to copy-draw one of six unique \textit{templates} as quickly and accurately as possible using an electronic stylus (\textit{One by Wacom\textregistered}, Wacom, Kazo, Japan; see Figure~\ref{fig:setup} A). Each template is a combination of three atoms. Per \textit{trial}, a template, defined in horizontal (x-direction) and vertical (y-direction) pixel coordinates, was displayed on a monitor at a comfortable viewing distance (approx. 70\,cm) while the cursor was controlled via a digital stylus. 
The cursor positions drawn across time, referred to as the \textit{trace} of a trial, were stored as tuples of x- and y-pixel coordinates, $\vec{\text{r}}_k=(\text{r}_k^x, \text{r}_k^y)$, where $k\in K=\{1,2,...,n_{\text{samples}}\}$ with $K$ being the index set of samples of a single drawing attempt (Figure~\ref{fig:setup} C). 
The experiment was conducted in \textit{blocks} of twelve trials each, with each unique template appearing twice but in random order within each block. DBS was alternated between blocks, resulting in DBS-ON and DBS-OFF blocks (Figure~\ref{fig:setup} B).

\subsection{Feature extraction}
\label{sec:met_features}
For each trial, numeric differentiation was applied to the measured trace (tuples of x- and y-pixel coordinates), using a convolutional smoothing with a sinc kernel on a modified Gaussian window (degree=6)~\cite{Schmid2022} with the window half-width fixed to 100\,ms. As the sampling rate differed between sessions (56--120\,Hz), the half-width was converted to 6--12 samples per session, resulting in a comparable frequency response across sessions ($-6$\,dB at 20--23\,Hz). Velocity ($v_k^x$, $v_k^y$), acceleration ($a_k^x$, $a_k^y$), and jerk ($j_k^x$, $j_k^y$) were derived successively for both the $x$ and $y$ directions, each by smoothing the preceding time series and computing its central difference quotient.
This approach produces smoother derivative curves and reduces the phase-shift compared to plain difference quotients and moving-average smoothing, which have been used before~\cite{Castano2019, Castano2020, dold2025, Dold2026}. To avoid edge artifacts, the first five samples (40--90\,ms, depending on the sampling rate) of each derivative time series were dropped.  

Out of a total of 5,363 trials, 32 had to be excluded after visual inspection due to the traces showing large streaks of interruption, originating from the stylus losing contact during the trial. In later experiments, the stump of the trace (until the first lost contact of the stylus) was considered to maintain as many trials as possible. In addition, 227 trials were excluded from all analyses, either because the recording software flagged them as fragmented (148 trials, only available for the newer recording system) or because they were marked as invalid in the experimental protocol (90 trials; 11 trials met both criteria). This left 5,104 trials (2,538 DBS OFF, 2,566 DBS ON) for the analyses.
\\
Alongside the direction-specific spatial features, the overall kinematic magnitudes ($\text{v}_k^m$, $\text{a}_k^m$, $\text{j}_k^m$) were calculated at each sample $k$ using the Euclidean norm (Figure~\ref{fig:setup} C).
\\
To quantify the drawing accuracy, a time-series of spatial distances $\text{d}_k := \sqrt{(\vec{r}_k - \vec{t}_j)^2}$ was calculated, where each trace point $\vec{r}_k$ was mapped to its spatially corresponding template point $\vec{t}_j$ using dynamic time warping (DTW)~\cite{Giorgino2009}. In some trials, the end of the template was reached before the time-out. To remove the meaningless end of such a trace, the trace was truncated after the first sample that matched the last point of the template. This led to an average reduction in the sample count ($n_{\text{samples}}$) per trial by $1.45\,\%$ (worst session $\text{S}14_4$ with $6.95\,\%$ reduction). 

Overall, the behavior observed in a trial was quantified by 38 scalar features, composed of three distinct feature subsets (Figure~\ref{fig:setup} C):

\textbf{Standard features (10):} These ten features correspond to the mean absolute values across the index set $K$ for the $x$-components, $y$-components, and magnitudes of velocity, acceleration, and jerk (e.g., $\text{v}_x = \langle |\text{v}_k^x| \rangle_K$, $\text{v}_y = \langle |\text{v}_k^y| \rangle_K$, $\text{v}_m = \langle |\text{v}_k^m| \rangle_K$, and analogously for $\text{a}_x,\text{a}_y$, $\text{a}_m$, $\text{j}_x$, $\text{j}_y$, $\text{j}_m$, $\text{d}$). With the exception of the distance (d), these are the features used to derive the CopyDraw scores in our previous work~\cite{Dold2026, dold2025}.

\textbf{Distributional features (12):} As distributional features we considered the standard deviation ($\sigma$), skewness (skew), and kurtosis (kurt) derived from the central moments of the time-series of four standard features ($\text{v}_k^m$, $\text{a}_k^m$, $\text{j}_k^m$ and $d_k$), quantifying the dispersion, asymmetry, and tailedness of the kinematic profiles. Including higher moments in the feature sets was motivated by a similar analysis performed on a spiral drawing task~\cite{Riosurrego2019}, where Rios-Urrego and colleagues found the kurtosis of the velocity highly informative for PD-stage classification. 
    
\textbf{Complexity features (16):} 
Hand-motor kinematics represent a downstream consequence of complex neural signals. The complexity of muscular and neurophysiological signals is known to be altered in PD~\cite{Averna2023}. It correlates with various disabilities~\cite{Hardstone2012, Li2019}, and has been proposed for PD-stage classification~\cite{Riosurrego2019} and tremor detection~\cite{Lopezdeipina2016}. Thus, we hypothesize that the non-linear dynamics of our drawing kinematics might also be modulated by DBS. To test this hypothesis, we computed four complexity features on the four kinematic time-series features ($\text{v}_k^m, \text{a}_k^m, \text{j}_k^m, \text{d}_k$). Each one assesses a distinct aspect of the signal's dynamics:

\begin{itemize}
    \item \textit{Sample Entropy} (SampEn)~\cite{Richman2000}: quantifies the overall regularity and predictability of the drawing kinematics; it has been shown to be altered in a finger-tapping task~\cite{McLinden2021};
    
    \item \textit{Permutation Entropy} (PermEn)~\cite{Bandt2002}: captures the local temporal ordering and dynamic volatility of the signal. It has recently been used to investigate skilled and non-skilled \handmotor tasks in PD~\cite{Ramos2025};
    
    \item \textit{Higuchi's Fractal Dimension} (HFD)~\cite{Higuchi1988}: assesses the structural roughness of the trace, which could capture micro-tremors; HFD was recently reported among the top three features for medication classification~\cite{Psathas2025}; 
    
    \item \textit{Detrended Fluctuation Analysis} (DFA)~\cite{Peng1994}: evaluates long-range temporal correlations indicative of fluid motor integration. It has been used as an indicator of various disabilities and motor dysfunctions~\cite{Hardstone2012, Li2019}.
\end{itemize}

These four complexity features were selected a priori to form a non-redundant description of the \handmotor dynamic time-series: While SampEn evaluates general, scale-dependent regularity, PermEn captures the ordinal structure of local patterns in a scale-free way. HFD complements these predictability features by quantifying the geometric complexity and scale-free structural roughness. Finally, whereas the previous three features capture localized, short-term dynamic properties, DFA assesses the long-range temporal (auto-)correlations.

\subsection{DBS effect on motor features}
\label{sec:met_dbs_effect}
In our previous studies~\cite{Castano2019, Castano2020, Dold2026} we investigated the impact of DBS on \handmotor kinematics by means of feature relevance in a machine learning (ML) based classification task. The ML-based approach has the benefit of ultimately extracting a single scalar score (CopyDraw score) that allows the identification of neural markers. However, it is limited with respect to understanding the DBS modulation of individual features for the following reasons: First, the ML models were fitted for each session individually. The resulting feature weights are session-specific and cannot be compared directly across sessions, as they depend on the session-specific regularization, scaling, and feature correlation structure. They therefore do not allow one to infer whether a feature is consistently modulated across sessions and patients, or whether its modulation differs between acute and chronic sessions. Second, while features might be informative in a (linear) combination for an ML model, this does not necessarily mean that they are significantly modulated in isolation. Finally, in a situation where features are collinear, feature importance can be shifted arbitrarily between such features, without changing the overall classification accuracy.

In this work, we therefore use linear mixed models (LMMs) to perform a detailed analysis of the DBS effect on \handmotor kinematics. LMMs are an established method for analyzing behavioral symptoms in PD~\cite{Lauro2023}, for quantifying the effects of medication and/or DBS~\cite{Giannini2025}, and for assessing the disease progression~\cite{Kamo2025}. For our data set, which consists of multiple sessions per subject, and multiple trials per session, LMMs provide the advantage of naturally accommodating the repeated measures design while capturing subject-specific variability in the DBS effect. To approximate a normal distribution and to stabilize the variance, all features were divided by their median absolute value, scaled using the Yeo-Johnson transformation~\cite{Yeo2000}, followed by standard z-scaling (subtracting the sample mean and dividing by the sample standard deviation). For the group-level LMMs, those steps were applied across all trials of all sessions, which preserves the differences between subjects and sessions that are modeled by the random effects. For the session-level models, those steps were applied per session. 

\subsubsection{Group effect of DBS on motor features}
To understand which features are modulated by DBS, we contrasted one LMM with DBS as independent variable (\textit{full model}) against an LMM without the DBS information (\textit{null model}). The models are defined in standard Wilkinson's notation\footnote{The left-hand side defines the dependent variable; the right-hand side defines the linear model with C($\cdot$) for categorical fixed effects. $\text{C(DBS)} * \text{C(session\_type)}$ resolves as C(DBS) + C(session\_type) + C(DBS):C(session\_type), i.e., the two independent fixed effects and the interaction term (C(DBS):C(session\_type)), which introduces a DBS effect that can be different for different session types (acute or chronic); ($1 \mid$ sub) is a random effect (intercept). (1 + C(DBS)$\mid$sub) resolves as ($1 \mid$ sub) and C(DBS)$\mid$sub, with the latter term describing a subject-specific random slope.} as:
\begin{align*}
    \textbf{full model:} \quad \text{feature} &\sim \text{C(DBS)} * \text{C(session\_type)} + \text{C(template\_id)} \\
    &\quad + \text{total\_ix\_trial} + (1 + \text{C(DBS)} \mid \text{sub}) + (1 \mid \text{sub\_ses}) \\[1.5ex]
    \textbf{null model:} \quad \text{feature} &\sim \text{C(session\_type)} + \text{C(template\_id)} \\
    &\quad + \text{total\_ix\_trial} + (1 \mid \text{sub}) + (1 \mid \text{sub\_ses})
\end{align*}

The model terms contained random effects for a subject specific intercept and slope for DBS (1 + C(DBS) $\mid$ sub) to capture the subject-specific response to DBS. A random intercept for each session accounts for session-specific baselines (1 $\mid$ sub\_ses). Fixed effects for the templates (identifier for the six unique templates, C(template\_id)) consider the different shapes, while a trial index term (total\_ix\_trial, index within all trials of a session) captures potential linear drifts across a full session, caused, e.g., by fatigue or other slow non-stationarities. 
A fixed effect for the session type (\textit{acute} in the first days post surgery, or \textit{chronic} after a few weeks to months post surgery, $\text{C(session\_type)}$) captures baseline differences between the two session types, while its interaction with DBS ($\text{C(DBS)}:\text{C(session\_type)}$) allows
the DBS effect to differ between acute and chronic sessions, e.g., due to a transient micro-lesioning effect shortly after implantation~\cite{Jech2012}.

We inferred a significant DBS effect on a feature when the full model yielded a significantly improved likelihood ($\mathcal{L}_\text{full}$) for the observed data relative to the null model ($\mathcal{L}_\text{null}$), as determined by a likelihood-ratio test ($\Lambda = -2 \log(\mathcal{L}_\text{null}/\mathcal{L}_\text{full})$, $\chi^2$ distribution, df=4).
Additionally, the Akaike information criterion (AIC)~\cite{Akaike1998} was calculated for both models and the difference $\Delta \text{AIC} =\text{AIC}_{\text{null}} - \text{AIC}_{\text{full}}$ was used to rank the features according to the DBS effect. 
The coefficient of the fixed DBS effect ($\beta_\text{stim}$, term C(DBS)) quantifies the average DBS effect across the cohort, whereas the standard deviation of the subject-specific random DBS slopes ($\text{SD}(\beta_{\text{stim}|\text{sub}})$, from the term $(1 + \text{C(DBS)} \mid \text{sub})$) quantifies how much this effect varies between subjects. Comparing both indicates whether the DBS effect is largely uniform ($\text{SD}(\beta_{\text{stim}|\text{sub}}) \ll |\beta_\text{stim}|$) or subject-specific ($\text{SD}(\beta_{\text{stim}|\text{sub}})$ comparable to or exceeding $|\beta_\text{stim}|$).

\subsubsection{Single session effect of DBS on motor features}
As we observed a high patient- and session-specific variance in the DBS effect across all features, we quantified the session-specific effect on individual features. First, we defined a \textit{nuisance model} for each feature, accounting for the template shape (template-id), potential slow drifts over the whole session (total trial number), and a within-block drift (relative trial number) covering potential wash-in/wash-out effects of the stimulation. The residuals of this nuisance model were then considered in a simple ordinary least squares (OLS) model (residuals $\sim$ C(DBS)). The block-clustered structure of the experiment was respected by using cluster-robust estimation of the standard errors (SE). In Wilkinson's notation, the models then read as:

\begin{align*}
    \textbf{nuisance model:} \quad \text{feature} &\sim \text{C(template\_id)} + \text{total\_ix\_trial} + \text{ix\_trial} \\[1.5ex]
    \textbf{residual model:} \quad \text{residual} &\sim \text{C(DBS)}
\end{align*}

In case of multiple testing, all reported $P$ values are corrected using the Benjamini-Hochberg method to control the false discovery rate (FDR). The correction is indicated by the explicit notation $P_{BH}$.

\subsection{DBS condition decoding from behavioral features}
\label{sec:met_decoding}
The CopyDraw task was designed to extract DBS-controllable behavioral scores for the identification of informative neural markers~\cite{Castano2020} and their final application in closed-loop DBS~\cite{dold2025}. In our recent work~\cite{Dold2026}, we observed that the neural decoding quality correlates with the behavioral classification accuracy measured by the area under the receiver operating characteristic curve (AUC). In other words, it is more likely to find a significant neural marker if the behavioral labels show a stronger DBS effect.

Based on this observation, we evaluate which input features yield the best behavioral decoding by applying the pipeline proposed in~\cite{Dold2026} to each of the three feature sets (\textit{standard}, \textit{distributional}, and \textit{complexity}) and to the union of all 38 features.
Compared to our previous work~\cite{Dold2026}, the scaling step of the pipeline was modified. The pipeline now contains dividing each feature by the median absolute value of the training set, followed by a Yeo-Johnson transformation~\cite{Yeo2000} before the z-scaling. 
The regularized~\cite{Ledoit2004} linear discriminant analysis (LDA) for the final classification was unchanged. All parameters of the pipeline were estimated on the training folds exclusively and then applied to the test fold.
The AUC was calculated as the mean across a chronological cross-validation, which paired chronologically adjacent blocks of different DBS states (ON, OFF) to create test sets, such that each block belonged to at most one test set (blocks could be excluded from the test sets if the number of DBS ON and OFF blocks did not match).
All other data remained in the training set. As the AUC was computed from the predicted class labels, it equals the balanced accuracy.
This approach was a trade-off between data efficiency and handling the chronological nature of the experiment session with potential non-stationarities impeding the generalizability of the decoding approaches. Chance levels were calculated as the 95\,\% quantiles of the AUC obtained by 1000 permutations with randomly shuffled labels in the training data.

%% file: results.tex
\section{Results}
The results are presented at three levels of granularity. We first assess which \handmotor features are modulated by DBS at the group level and how consistent this effect is across subjects (Section~\ref{sec:res_group}), then examine the DBS effect within individual sessions (Section~\ref{sec:res_session}), and finally compare the four feature sets in their ability to decode the DBS state (Section~\ref{sec:res_decoding}).

\subsection{Group level analysis}
\label{sec:res_group}
\begin{figure}
    \centering   
    \includegraphics[width=\linewidth]{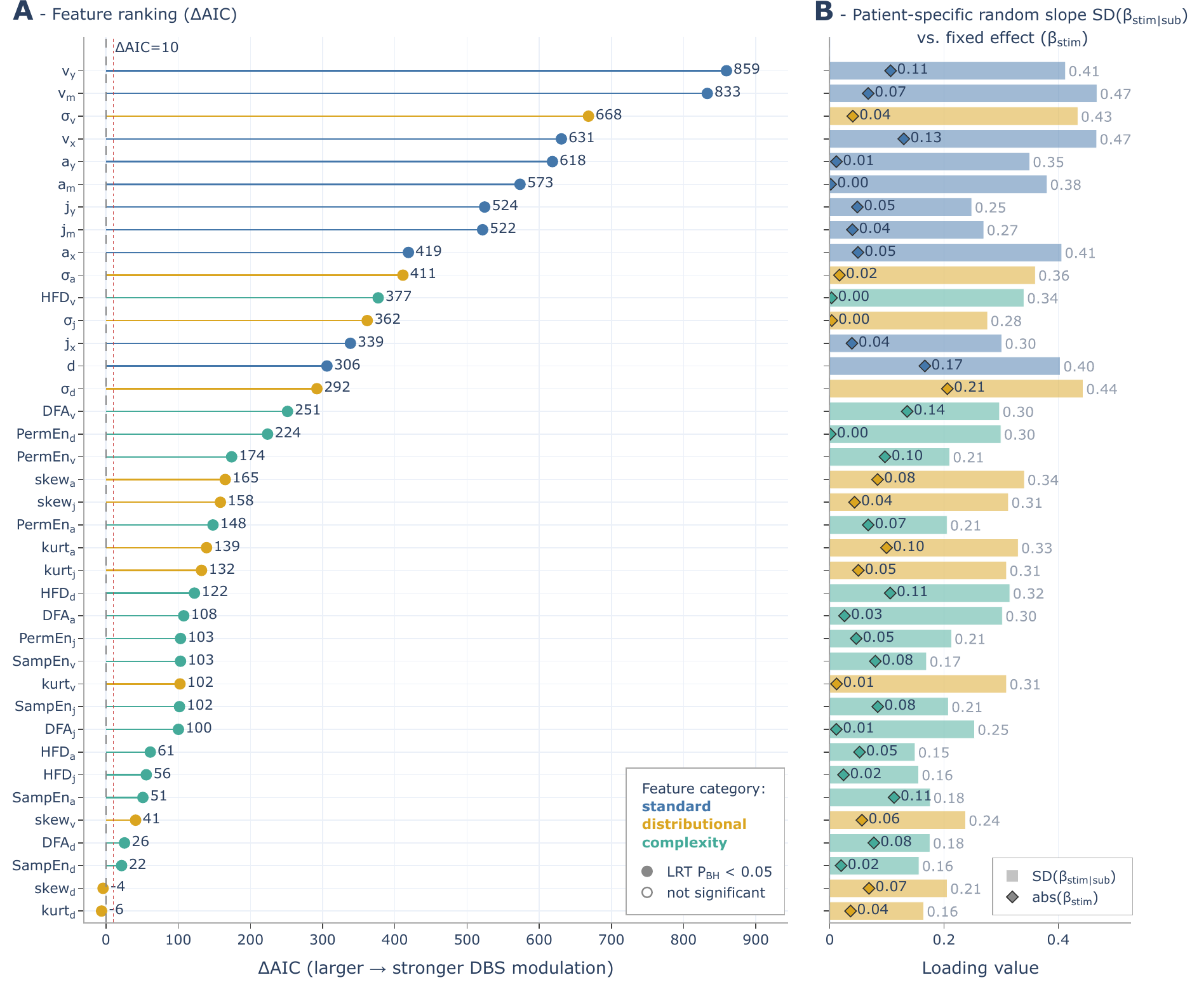}      
    \caption{Group level analysis of the DBS impact, using a global linear mixed model. \textbf{A}: Features sorted by difference in the Akaike information criterion ($\Delta$AIC) between the full model (including the DBS terms) and the null model (without DBS terms). The colors indicate the feature categories (\textit{standard} in blue, \textit{distributional} in orange and \textit{complexity} in green). A dashed line is drawn for $\Delta\text{AIC}=10$, a rule of thumb criterion for a strong difference. The circle markers are filled for significant, FDR-corrected \textit{P} values. \textbf{B}: Group level fixed effect absolute coefficient ($\text{abs}(\beta_\text{stim})$) and standard deviation for the random slope ($\text{SD}(\beta_{\text{stim}|\text{sub}})$) of the DBS terms in the full model. The y-axis and the color coding are shared with panel A.
    }
    \label{fig:group_level}
\end{figure}

To evaluate which \handmotor features were modulated by DBS across the population, we compared an LMM containing the DBS state (full model) against a null model without the DBS state information. The difference in the Akaike Information Criterion ($\Delta$AIC) between these models, alongside likelihood ratio tests (LRT), was used to rank the features by their sensitivity to DBS and to assess whether including the DBS state led to a significant model improvement (p-value of the LRT).
The group level analysis is presented in Figure~\ref{fig:group_level}. Panel A shows the $\Delta$AIC between the full and the null model for each of the 38 features, ranked by the size of the difference. The top ranks are mainly filled by features of the \textit{standard} kinematics, and overall predominantly by velocity features. All 38 features reach significance in the LRT. For the kurtosis ($\text{kurt}_\text{d}$) and skew ($\text{skew}_\text{d}$) of the distances between the drawn trace and the template, however, the improvement is marginal ($P_{BH}=0.044$ and $P_{BH}=0.022$) and the $\Delta$AIC is negative, i.e., the AIC favors the null model.

While a DBS effect on the group level is observed for all features, it is highly individual. Panel B illustrates the absolute value of the coefficient of the fixed DBS effect ($\text{abs}(\beta_\text{stim})$) vs.~the standard deviation of the random slope coefficient ($\text{SD}(\beta_{\text{stim}|\text{sub}})$). This serves as a comparison of the group level homogeneous DBS effect against the subject-specific one. The figure shows that $\text{abs}(\beta_\text{stim})$ can be very small while $\text{SD}(\beta_{\text{stim}|\text{sub}})$ reaches values of up to $0.47$ (i.e., $47\,\%$ of the feature's standard deviation, as the features were z-scaled before model fit), overshadowing the mean fixed effect coefficient substantially. 

\subsection{Session level analysis}
\label{sec:res_session}
\begin{figure}
    \centering   
    \includegraphics[width=\linewidth]{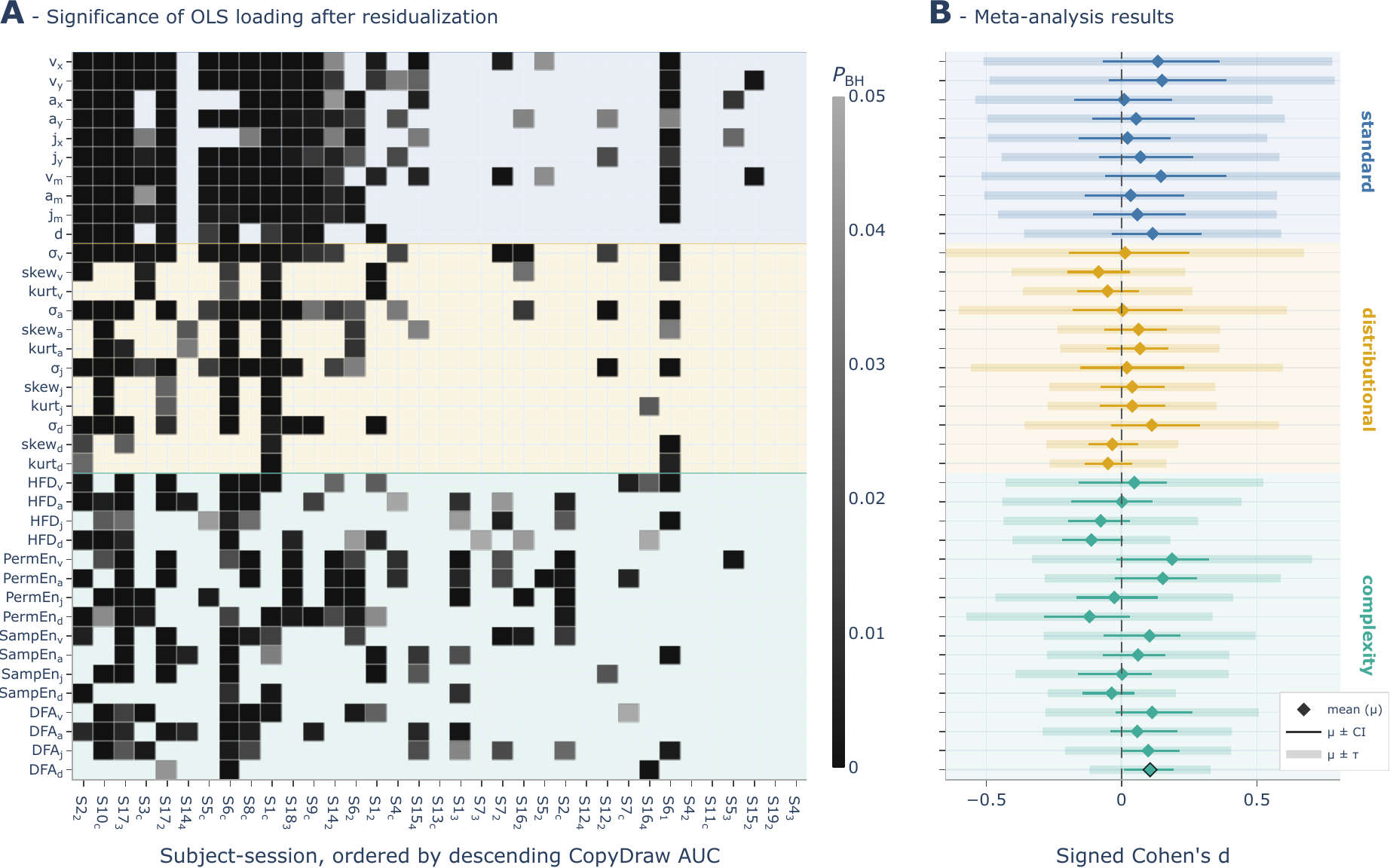}      
    \caption{Session-specific DBS impact on features: session-specific models. \textbf{A}: Heatmap of \pval for the DBS effect on residualized features. Residuals are obtained per session by regressing each feature against the template shape (template\_id), total trial number and within block trial number (feature $\sim$ C(template\_id) + total\_ix\_trial + ix\_trial) for each session separately, using ordinary least-squares regression (OLS). The DBS effect is then estimated on the residuals via OLS (residuals $\sim$ C(DBS)) with cluster-robust standard errors grouped by recording blocks (ix\_block). P-values are corrected for multiple comparisons across sessions. Only cells with \pval\,<\,0.05 are shown; sessions are ordered along the x-axis by descending CopyDraw AUC as reported by Dold et al. 2026~\cite{Dold2026}, and features (y-axis in both A and B) are grouped by category (standard, distributional, complexity). \textbf{B}: Random-effects meta-analytic forest plot of Cohen's d across sessions (x-axis) for each feature (y-axis, labels as in A). The point estimate (diamond) represents the random-effects mean effect ($\mu$), error bars span the 95\,\% confidence interval of $\mu$. The wider shaded band ($\mu\pm\tau$) illustrates the between-session standard deviation of the stimulation effect. Outlined diamonds mark features for which the population mean estimate is significant ($\mu\pm\text{CI}$ excludes zero).
    }
    \label{fig:session_level}
\end{figure}

To better understand the session-specific DBS effect, we used a nuisance model to remove shape-specific and linear temporal non-stationarities. We then evaluated the coefficients of a standard ordinary least-squares (OLS) model fitted on the residuals.
Figure~\ref{fig:session_level} illustrates the results of the residualized model per session. Panel A displays the feature cross-session matrix, with cells shaded in gray if the FDR-corrected P-value of the coefficient of the residual model (residuals $\sim$ C(DBS)) is significant (\pval\,<\,0.05). The sessions on the x-axis are sorted according to how well the DBS state was decoded from motor features (the CopyDraw AUC) reported in~\cite{Dold2026}. Sessions with a higher AUC are on the left. Session names are composed of the subject identifier (S1--S19) and a subscript. The subscript denotes the session: one to four ($\cdot_1$, $\cdot_2$, $\cdot_3$, $\cdot_4$) indicates the number of days post surgery for acute sessions, and $\cdot_c$ indicates chronic sessions.
Notably, sessions with a higher AUC show a more frequent significant modulation of features (more filled cells towards the left). Standard features and complexity features are modulated more homogeneously than distributional features. Multiple sessions ($\text{S}1_3$, $\text{S}7_2$, $\text{S}16_2$, $\text{S}2_c$) show a sparse modulation in the standard features but have multiple entropy features (PermEn and SampEn) with significant modulation by DBS. To generalize across sessions, a meta-analysis approach is shown in Panel B. The plot visualizes the mean ($\mu$), 95\,\% confidence interval (CI), and the between-session standard deviation ($\tau$) of Cohen's d derived for each feature. The range $\mu\pm\tau$ is relatively large and indicates that the magnitudes and directions of the DBS effect are highly variable across sessions.

Despite this widespread heterogeneity, only the detrended fluctuation analysis (DFA) of the distances ($\text{DFA}_\text{d}$) demonstrates a statistically significant, non-zero population mean effect ($\mu=0.11$, 95\,\% CI $[0.01, 0.19]$; black outlined marker in Figure~\ref{fig:session_level} B). The DFA of the velocity ($\text{DFA}_\text{v}$, $\mu=0.11$, 95\,\% CI $[-0.02, 0.26]$), the permutation entropy of the velocity ($\text{PermEn}_\text{v}$, $\mu=0.19$, 95\,\% CI $[-0.02, 0.32]$), and the skewness of the velocity ($\text{skew}_\text{v}$, $\mu=-0.09$, 95\,\% CI $[-0.20, 0.03]$) miss significance, but show a trend.

\subsection{DBS condition decoding from motor features}
\label{sec:res_decoding}
We evaluated machine learning (ML) pipelines for classifying DBS ON vs.~OFF based on the three feature sets (standard, distributional, complexity) as well as a combined feature set including all 38 features. Mean AUC values per feature set, derived from chronological cross-validation within each session, are reported in Table~\ref{tab:roc_per_session_feature_type}.  Overall, the combined feature set provides the highest grand average performance ($\text{AUC}=0.700$, standard: $\text{AUC}=0.677$) and is the optimal choice in 13 out of 35 sessions. Notably, the standard feature set is the best choice for ten out of 35 sessions. The distributional or complexity sets alone are optimal choices in six sessions each.
An additional column in Table~\ref{tab:roc_per_session_feature_type} displays the \textit{Oracle} AUC, i.e., choosing the best feature set a posteriori. This is equivalent to how a decoding model would be chosen for a downstream application in identifying controllable neural markers~\cite{Castano2020}, but introduces a feature selection bias when considering the AUC in isolation.

\begin{table}[ht]
\centering
\small
\caption{Mean ROC AUC per session and feature set (LDA, division by the median absolute value, Yeo-Johnson and z-scaling, chronological CV). Bold: best feature set per session. Oracle: cherry-picked best AUC per session (upper bound from selecting the optimal feature set in hindsight). $^\dagger$: not significant, i.e. the AUC does not exceed the 95th percentile of the per-session label shuffle distribution (1000 shuffles of all training trial labels).}
\label{tab:roc_per_session_feature_type}
\begin{tabular}{lrrrrr}
\toprule
    & \multicolumn{4}{c}{\textbf{Feature sets}} & \\
\cmidrule(lr){2-5}
    \textbf{Session} & \textbf{Standard} & \textbf{Distributional} & \textbf{Complexity} & \textbf{Combined} & \textbf{Oracle} \\
\midrule
    S10$_{c}$ & $0.958$ & $0.847$ & $0.958$ & $\mathbf{0.972}$ & $\mathbf{0.972}$ \\
    S2$_{2}$ & $\mathbf{0.938}$ & $0.924$ & $0.847$ & $0.931$ & $\mathbf{0.938}$ \\
    S17$_{3}$ & $0.933$ & $\mathbf{0.935}$ & $0.894$ & $0.925$ & $\mathbf{0.935}$ \\
    S6$_{c}$ & $\mathbf{0.870}$ & $0.828$ & $0.802$ & $0.865$ & $\mathbf{0.870}$ \\
    S8$_{c}$ & $\mathbf{0.833}$ & $0.786$ & $0.786$ & $0.798$ & $\mathbf{0.833}$ \\
    S3$_{c}$ & $0.812$ & $0.766$ & $0.816$ & $\mathbf{0.824}$ & $\mathbf{0.824}$ \\
    S17$_{2}$ & $0.789$ & $0.810$ & $\mathbf{0.839}$ & $0.838$ & $\mathbf{0.839}$ \\
    S5$_{c}$ & $0.776$ & $0.635$ & $0.604$ & $\mathbf{0.797}$ & $\mathbf{0.797}$ \\
    S1$_{c}$ & $\mathbf{0.776}$ & $0.754$ & $0.697$ & $0.730$ & $\mathbf{0.776}$ \\
    S6$_{2}$ & $0.757$ & $0.757$ & $0.736$ & $\mathbf{0.764}$ & $\mathbf{0.764}$ \\
    S18$_{3}$ & $0.738$ & $\mathbf{0.743}$ & $0.703$ & $0.737$ & $\mathbf{0.743}$ \\
    S9$_{c}$ & $0.721$ & $0.638$ & $0.656$ & $\mathbf{0.750}$ & $\mathbf{0.750}$ \\
    S14$_{2}$ & $0.720$ & $0.672$ & $\mathbf{0.741}$ & $0.707$ & $\mathbf{0.741}$ \\
    S1$_{2}$ & $0.711$ & $\mathbf{0.713}$ & $0.640$ & $0.700$ & $\mathbf{0.713}$ \\
    S6$_{1}$ & $0.708$ & $\mathbf{0.771}$ & $0.667$ & $0.729$ & $\mathbf{0.771}$ \\
    S14$_{4}$ & $0.675$ & $0.603$ & $0.734$ & $\mathbf{0.747}$ & $\mathbf{0.747}$ \\
    S13$_{c}$ & $\mathbf{0.667}$ & $0.658$ & $0.500^\dagger$ & $0.658$ & $\mathbf{0.667}$ \\
    S4$_{c}$ & $\mathbf{0.646}$ & $0.597$ & $0.604$ & $0.618$ & $\mathbf{0.646}$ \\
    S5$_{2}$ & $0.632$ & $0.444^\dagger$ & $\mathbf{0.667}$ & $0.667$ & $\mathbf{0.667}$ \\
    S2$_{c}$ & $0.622$ & $0.491^\dagger$ & $\mathbf{0.706}$ & $0.699$ & $\mathbf{0.706}$ \\
    S12$_{2}$ & $\mathbf{0.622}$ & $0.559^\dagger$ & $0.517^\dagger$ & $0.538^\dagger$ & $\mathbf{0.622}$ \\
    S16$_{4}$ & $\mathbf{0.620}$ & $0.583^\dagger$ & $0.607$ & $0.615$ & $\mathbf{0.620}$ \\
    S15$_{4}$ & $0.610$ & $0.531^\dagger$ & $0.652$ & $\mathbf{0.692}$ & $\mathbf{0.692}$ \\
    S16$_{2}$ & $0.603$ & $0.564^\dagger$ & $0.541^\dagger$ & $\mathbf{0.627}$ & $\mathbf{0.627}$ \\
    S7$_{2}$ & $0.603$ & $0.695$ & $0.682$ & $\mathbf{0.779}$ & $\mathbf{0.779}$ \\
    S7$_{c}$ & $0.599$ & $0.573$ & $\mathbf{0.656}$ & $0.641$ & $\mathbf{0.656}$ \\
    S15$_{2}$ & $\mathbf{0.599}$ & $0.455^\dagger$ & $0.523^\dagger$ & $0.570^\dagger$ & $\mathbf{0.599}$ \\
    S7$_{3}$ & $0.595$ & $0.554^\dagger$ & $0.619$ & $\mathbf{0.696}$ & $\mathbf{0.696}$ \\
    S5$_{3}$ & $0.569^\dagger$ & $0.472^\dagger$ & $0.569^\dagger$ & $\mathbf{0.681}$ & $\mathbf{0.681}$ \\
    S12$_{4}$ & $\mathbf{0.567}^\dagger$ & $0.425^\dagger$ & $0.442^\dagger$ & $0.500^\dagger$ & $\mathbf{0.567}^\dagger$ \\
    S1$_{3}$ & $0.552^\dagger$ & $0.546^\dagger$ & $\mathbf{0.656}$ & $0.646$ & $\mathbf{0.656}$ \\
    S11$_{c}$ & $0.529^\dagger$ & $0.523^\dagger$ & $0.497^\dagger$ & $\mathbf{0.556}^\dagger$ & $\mathbf{0.556}^\dagger$ \\
    S4$_{2}$ & $0.500^\dagger$ & $\mathbf{0.548}^\dagger$ & $0.488^\dagger$ & $0.530^\dagger$ & $\mathbf{0.548}^\dagger$ \\
    S19$_{2}$ & $0.467^\dagger$ & $\mathbf{0.496}^\dagger$ & $0.450^\dagger$ & $0.477^\dagger$ & $\mathbf{0.496}^\dagger$ \\
    S4$_{3}$ & $0.369^\dagger$ & $0.464^\dagger$ & $0.476^\dagger$ & $\mathbf{0.482}^\dagger$ & $\mathbf{0.482}^\dagger$ \\
\midrule
    \textit{Grand mean} & $0.677$ & $0.639$ & $0.656$ & $\mathbf{0.700}$ & $\mathbf{0.714}$ \\
\bottomrule
\end{tabular}
\end{table}

%% file: discussion.tex
\section{Discussion}

This work investigates the effect of DBS on \handmotor features for 35 sessions of 19 patients with Parkinson's disease performing the CopyDraw task. Linear mixed models (LMM) and regression analysis are used to discern DBS-induced differences in velocity, acceleration, jerk, and accuracy of copy drawing target templates using a digital stylus. In addition to this \textit{standard} set of mean features, which was analyzed in a machine learning feature importance context before~\cite{Castano2019}, two new feature sets including \textit{distributional} higher moments, i.e., standard deviation, skew, and kurtosis, as well as \textit{complexity} features, i.e., sample entropy, permutation entropy, Higuchi's fractal dimension, and detrended fluctuation analysis, were analyzed. The analysis reveals that while standard features provide a robust a priori choice for capturing DBS effects, the effect remains highly session-specific. Distributional and complexity features are found to be significantly modulated as well and provide complementary information. 

\subsection{Group level analysis}
The results of the group level analysis comparing the full LMM (with DBS terms) and the null model (without DBS terms) are depicted in Figure~\ref{fig:group_level} A. This figure reveals that for all but two features ($\text{kurt}_\text{d}$ and $\text{skew}_\text{d}$), a model with DBS terms describes the observed data significantly better (when considering the $\Delta$AIC in additiona to the LRT). Such an effect is to be expected, as PD is known to lead to deficits in handwriting~\cite{Thomas2017, Letanneux2014} as a result of impaired basal ganglia-thalamocortical loops. 

The amount of fit improvement can be quantified by the difference in AIC ($\Delta$AIC). Here, we observed a stronger difference for the \textit{standard} feature set category, as compared to the distributional and complexity features. The \textit{standard} feature set, containing the per-trial mean velocity, acceleration, jerk and distance between drawn trace and target template, would hence be the preferred prior for an ML-based decoding setup.

Furthermore, features derived from the velocity measures ($\text{v}_{x}$, $\text{v}_{y}$, $\text{v}_{m}$) show an overall stronger difference compared to those derived from acceleration, jerk, or the distance measure.
Finding velocity amongst the top features is expected, as velocity in handwriting and drawing is linked to bradykinesia, a hallmark symptom of PD~\cite{Bloem2021}. Besides velocity features, we also report acceleration features among the top eight with the highest $\Delta$AIC. Again, this observation is consistent with literature reporting central moments of both velocity and acceleration being strong predictors for classifying healthy controls from patients with PD in \handmotor tasks~\cite{Smits2014}. 

Conversely, we observed less modulation for the jerk features by DBS, compared to their velocity and acceleration counterparts (Figure~\ref{fig:group_level} A). We hypothesize that the jerk features would be mainly driven by tremor symptoms, which were very mild in the participant population. In PD overall, tremor is usually dominant in the resting subtype~\cite{Hauser2025} and might therefore not manifest itself during the continuous and goal-directed movement of the CopyDraw task. Another possible explanation arises from the different origin of tremor stemming from a cerebello-thalamo-cortical system~\cite{Abusrair2022}. This system was shown not to depend on the volume/activity of substantia nigra (SN)~\cite{Poston2020} as strongly as, e.g., bradykinesia and rigidity. Together with the observation that bradykinesia and tremor periods are distinct and anticorrelated during handwriting~\cite{Lauro2023}, it seems likely that the CopyDraw task is sensitive to the modulation of the bradykinesia subsystem, but not the tremor one.  

Regarding the distance measure, we note that PD patients can maintain accuracy in \handmotor tasks~\cite{Fernandez2018, Mazzoni2007} similar to healthy controls. This could explain why features derived from the distance measure (or accuracy as the inverse of distance) are less modulated by DBS than the purely kinematic ones. 

We found that the ranking of these different measures is consistent with the feature importance analysis of session-specific machine learning classifiers on the same data set~\cite{Dold2026}. Shapley additive explanation values~\cite{Lundberg2017} attributed most importance to the velocity features---predominantly for sessions with a high classification accuracy. This points to an important limitation when only the group-level effect is reported, as it might be dominated by single individuals. This bias is especially strong when the sample size is small, as is commonly the case for DBS-related research~\cite{Hu2025}.

The difference between group-level means and subject-specific effects is clearly visible in Figure~\ref{fig:group_level} B. Large standard deviations of the random slope are observed with comparatively small values of the fixed effect (group wide) DBS coefficient. Similar differences between group-level and subject-specific effects have been reported in larger multi-center studies~\cite{Gerster2025}. Such findings go beyond behavioral or symptomatic observations and concern neural signatures, such as beta-band power of local field potentials, for which different correlations with symptom severity are observed in the group vs.~the individual level~\cite{Gerster2025}. This high inter-subject variability clearly mandates individualized analysis approaches when dealing with DBS.

\subsection{Session-level analysis}
From previous ML-based analyses of the data set~\cite{Dold2026} we expected session-specific differences in the DBS effect on behavioral features. Our session-specific analysis of residualized features, after removing the influence of shape and linear drifts, confirms this observation, as visualized in the heatmap of Figure~\ref{fig:session_level} A. 
Sessions with a strong ML-based classification (AUC) reveal consistent differences in the residualized features for both the standard and the complexity feature sets. While this raises the question of how much information is redundant between these two sets, we also observed multiple sessions ($\text{S}1_3$, $\text{S}7_2$, $\text{S}16_2$, $\text{S}2_c$) for which complexity features are more often significantly modulated than standard features, i.e., sessions for which complexity features provide complementary information. Here, DBS seems to modulate drawing capabilities which are not captured in the statistical moments of the feature distributions (standard and distributional feature sets).

The meta-analysis of the per-session effect sizes (forest plot of Cohen's d, Figure~\ref{fig:session_level} B) reveals a picture similar to the group analysis: The average effect is small ($\mu$), but the between-session variability ($\tau$) is large. In particular, the range $\mu\pm\tau$ spans both signs. Hence, DBS might improve a particular motor feature for one patient with PD, while it worsens the same feature for another patient. Only one feature has a significant effect in the meta-analysis: the detrended fluctuation analysis of the distances ($\text{DFA}_\text{d}$).

DFA measures the strength of long-lasting correlations. Based on accelerometric data obtained from an elderly cohort, DFA was shown to be a predictor for frailty, disability, and mortality~\cite{Li2019}. Higher DFA values were associated with a healthier condition---an observation shared with various other neurological conditions~\cite{Hardstone2012}. Assuming a general motor-symptom improvement under DBS, our findings regarding DFA point in the same direction.

Three other features showed a non-significant trend. The DFA of the velocity was higher under DBS ON, in the same direction as $\text{DFA}_\text{d}$. Consistent with previous findings, the velocity distribution was less skewed and the permutation entropy (PermEn) of velocity was higher under DBS ON.

Patients with PD are expected to show a positively skewed velocity distribution (overall slower movements, with a few very fast movements) compared to more normally distributed velocities obtained from healthy controls~\cite{Shin2025}. The observed trend is in line with an expected therapeutic improvement.

PermEn captures the richness of local, ordinal patterns in a time series. An increase in permutation entropy due to DBS might therefore reflect a restoration of flexibility of an otherwise rigid system, bound to less versatile movement patterns. A recent study analyzed in parallel the PermEn from handwriting and drawing tasks, and the PermEn obtained from electroencephalography signals~\cite{Ramos2025}. The authors established a very interesting link between the neural control and the resulting motor execution with a unified complexity metric. They further show that PermEn in the behavioral data is higher for the more skilled task of writing as opposed to drawing, and that it is higher in the dominant hand. A higher PermEn might thus reflect a more automated neural control mode, which potentially is restored by DBS. 

Future work on the data set might explore the potentially similar complexity characteristics of biomechanical and neural time series~\cite{Hardstone2012, Ramos2025}.

\subsection{DBS condition decoding from motor features}
Considering the ML decoding pipelines with additional features leads back to the original purpose of the CopyDraw task, which is to provide a quick and objective assessment of \handmotor performance that is maximally modulated by DBS~\cite{Castano2019}. Considering the three individual feature sets in Table~\ref{tab:roc_per_session_feature_type}, we observe that overall, the per-session analysis using residualized regression and the ML approach both provide a similar view: A high AUC based on the standard feature set, i.e., a strong difference under DBS, usually implies a high AUC in the distributional and complexity sets as well. The different feature sets partly carry redundant information with respect to the classification task. A combined model, receiving all 38 features, outperformed the individual feature sets in 13 out of 35 sessions. A benefit of including more features in the model mainly manifests in sessions that suffer from an overall lower AUC. This reflects non-redundant information of features beyond the standard feature set or the model's capability to resolve more subtle changes that are encoded only in the combination of features. However, it also increases the risk of overfitting, which even our chronological cross-validation can only partly mitigate. 
Overall, the standard feature set remains a strong a priori choice and provided the highest AUC in ten out of 35 sessions. A richer feature vector should be considered if the standard features are not leading to a sufficient AUC.

\subsection{Limitations and future work}
Our study provides novel insights into the DBS modulation of \handmotor features of the CopyDraw task, but it is limited by a number of factors. First, the additional distributional and complexity features only represent a first selection as motivated in the Methods section. Multiple other features, especially from the complexity category, could be investigated but would require a potentially extensive feature selection~\cite{Psathas2025}. Given the limited amount of available data and an inherent risk of overfitting, we decided against expanding the feature sets further. 

Second, our evaluation did not discriminate between therapeutic responders and non-responders. While response rates to DBS are usually high (>70\,\%)~\cite{CastilloTorres2025, Morgante2025}, reasons for non-response are varied~\cite{Rossi2018} and aspects like the micro-lesioning effect~\cite{Jech2012} have to be considered for the acute sessions. It is possible that a clearer and more interpretable group level analysis would result from considering responders only.

Finally, the restrictions related to the data set, as outlined in~\cite{Dold2026}, apply here as well: A longitudinal evaluation of the DBS effect was not part of the data set, such that the stability of the effect across multiple days/weeks could not be explored. In addition, no systematic medication control was applied. Lastly, the CopyDraw task is not standardized and has not been systematically compared to, e.g., the MDS-UPDRS~\cite{Goetz2008}, which due to its nature is subjective, discrete, and ordinal, having its own limitations. 

As a consideration for future work we propose using (pen) pressure values from the digitizer, which have been shown to deliver relevant features in multiple studies classifying PD stages~\cite{Zham2017} or comparing patients with PD and healthy controls~\cite{Riosurrego2019, Drotar2016}. No pressure values were available in the presented data set.

%% file: conclusion.tex
\section{Conclusion}
In conclusion, this study demonstrates that DBS significantly modulates several categories of \handmotor kinematics during the CopyDraw task, with the nature and magnitude of this modulation being highly patient- and session-specific.
At the group level, standard kinematic features---particularly mean velocity---exhibit the strongest DBS-induced modulation.
At the single-session level, distributional and complexity features provide additional and complementary information with respect to DBS state.
These features suggest that DBS also affects the variability and temporal structure of \handmotor control, most consistently the long-range temporal correlations of the drawing error (DFA of the distance to the template).
Finally, while standard features remain a good a priori choice for behavioral decoding, integrating complexity and distributional features can enhance decoding performance in otherwise challenging sessions, and could serve as more sensitive behavioral observations for the identification of neural markers for potential closed-loop DBS. Future work should focus on optimizing subject-specific decoding approaches, evaluating feature stability across sessions, and implementing these models in real-time frameworks for adaptive DBS.

%

%% file: acknowledgements.tex
\section{Acknowledgments}

MD receives funding from the Dareplane collaboration project, which is co-funded by PPP Allowance awarded by Health~Holland, Top Sector Life Sciences \& Health, to stimulate public-private partnerships as well as by a contribution from the Dutch Brain Foundation. The PD-Interaktiv I study is funded by the Bundesministerium für Bildung und Forschung (Grant 16SV8011). 
VAC receives a collaborative grant from BrainLab (Munich, Germany). He serves as an advisor for Aleva (Lausanne, Switzerland), Ceregate (Hamburg, Germany), Cortec (Freiburg, Germany) and Inbrain (Barcelona, Spain). He has an ongoing IIT with Boston Scientific (USA). He has received travel support and honoraria for lectures from Boston Scientific (USA), UNEEG Medical (Munich, Germany), and Precisis (Heidelberg, Germany). The authors thank the EuroDBS Alliance (EuroDBS.org) for feedback and discussions on this work. 
BS is funded by the Berta-Ottenstein-Programme for Advanced Clinician Scientists, Faculty of Medicine, University of Freiburg. He serves as a scientific advisor for Precisis (Heidelberg, Germany) and has received a research grant from Ceregate (Hamburg, Germany) and travel support from Boston Scientific (Marlborough, MA, USA) and honoraria as a speaker and travel support from Medtronic PLC (Dublin, Ireland) and Insightec (Tirat Carmel, Israel).
JP is funded by the Deutsche Forschungsgemeinschaft (DFG) under the Walter Benjamin Program (Project number 510112977). The other authors declare no conflicts of interest.

\section{Author Contributions}
All authors: Writing - Review and Editing. MD: Conceptualization, Methodology, Software, Validation, Formal Analysis, Investigation, Data Curation, Writing - Original Draft, Visualization, Project Administration.  JP: Conceptualization, Methodology, Investigation, Project Administration, Writing - Original Draft. VAC: Conceptualization, Methodology, Resources, Supervision, Project Administration, Funding Acquisition. BS: Investigation, Project Administration. MR: Visualization. SG: Investigation, Data Curation. YT: Resources, Supervision, Funding Acquisition. MLFJ: Conceptualization, Investigation, Resources, Writing - Original Draft, Supervision, Project Administration, Funding Acquisition. MT: Conceptualization, Methodology, Investigation, Resources, Writing - Original Draft, Supervision, Project Administration, Funding Acquisition.